# An Efficient Two-Port Electron Beam Splitter via Quantum Interaction-Free Measurement

Yujia Yang,[1] Chung-Soo Kim,[1] Richard G. Hobbs,[1] Pieter Kruit,[2] Karl K. Berggren[1*]

[1]*Research Laboratory of Electronics, Massachusetts Institute of Technology, Cambridge, Massachusetts 02139, United States*

[2]*Department of Imaging Physics, Delft University of Technology, Lorentzweg 1, 2628CJ Delft, The Netherlands*

Semi-transparent mirrors are standard elements in light optics for splitting light beams or creating two versions of the same image. Such mirrors do not exist in electron optics, although they could be beneficial in existing techniques such as electron interferometry and holography and enable novel electron imaging and spectroscopy techniques. We propose a design for an electron beam splitter using the concept of quantum interaction-free measurement (IFM). The design combines an electron resonator with a weak phase grating. Fast switching gates allow electrons to enter and exit the resonator. While in the resonator, the phase grating transfers intensity from the direct beam into one of the weakly diffracted beams at each pass. To make the beam splitter an efficient two-port splitter, the intensity in all other diffracted beams is blocked by an aperture. The IFM principle minimizes the loss of total intensity by this aperture. We use a scattering matrix method to analyze the performance of the beam splitter, including the effects of inelastic scattering in the phase grating. This design can be generalized to beam splitters for not only electrons, but also photons, neutrons, atoms, and other quantum mechanical systems.

---

[*] berggren@mit.edu

## I. INTRODUCTION

Electron beam splitters are used in many applications such as electron interferometry [1], holography [2], imaging [3], and spectroscopy [4,5]. These applications benefit from the short de Broglie wavelength of electrons and a strong electron-matter interaction. However, many of these applications require a coherent and efficient two-port beam splitter, which cannot be readily provided by existing electron beam splitters. In light optics, efficient two-port beam-splitting can be achieved by using either a half-silvered mirror, a waveguide coupler, or a fiber switch. Unfortunately, all of these techniques are difficult to implement for electron beams. In this work, we propose a two-port electron beam splitter that uses quantum interference to realize near-ideal efficiency.

Several types of electron beam splitters have been developed previously: biprisms [6], crystals [7], optical standing waves [8], and nanofabricated gratings [5,9–13]. The most commonly used electron beam splitters are biprisms, which split the incoming electron beam into two output beams by the electrostatic force of a charged wire. A biprism placed in the electron beam inevitably blocks a certain portion of the beam and causes diffraction effects due to the wire edges, leading to intensity loss. Additionally, a biprism is a wavefront-division beam splitter, which divides the wavefront of the incoming electron beam. Wavefront-division beam splitters usually require a highly coherent point electron source [14]. This class of beam splitters also cannot split a wave with a pattern in it and hence cannot be easily applied to emerging electron beam technologies that use quantum mechanical effects [3,15]. Alternatively, thin crystals have been used as amplitude-division electron beam splitters. These beam splitters have a less stringent requirement on illumination coherence, leading to higher intensities by using extended sources, and a larger interference field [14]. However, electron diffraction from a crystal typically results in multiple diffracted beams. In order to make a two-port beam splitter, high-order diffracted beams need to be blocked, which leads to intensity loss. Electron beams can also be diffracted with optical standing waves by using the Kapitza-Dirac (KD) effect [8]. The advantage of KD effect is that electrons do not need to go through or near materials, thus minimizing decoherence caused by inelastic scattering. However, the KD effect requires high quality laser beams and good alignment, and it still suffers from finite intensity loss due to high-order diffraction. Recently, nanofabricated gratings have been proposed as electron beam splitters [9–12]. These beam splitters are also amplitude-division beam splitters, and has a less stringent requirement on the coherence condition of illumination. Nanofabrication also enables the production of arbitrary patterns to modulate the incoming electron beam, inspiring new applications such as electron vortex beam generation [5,11,12,16]. For the typical electron energies used in an electron microscope, thin membrane nanofabricated gratings are fairly transparent and there is only a small intensity loss due to inelastic scattering. For example, a 30–nm-thick silicon nitride membrane inelastically scatters roughly 20% of a 300 keV incident electron beam [12]. Thinner membranes result in even less intensity loss. However, similar to crystal beam splitters, intensity loss due to high-order diffraction remains as a problem for nanofabricated gratings. For example, a grating patterned into a 30-nm-thick silicon nitride membrane shows ~34% maximum first order diffraction efficiency (with respect to the transmitted beam) [12].

There have been some attempts to increase the efficiency of electron beam splitters. For biprisms, selecting a wire with a small diameter helps to reduce intensity loss [10]. For crystals and optical standing waves, a "two-beam" condition, or Bragg regime, can be achieved by tilting the beam splitters so that only one diffracted beam is strongly excited. However, even in this situation, there is still finite intensity in high-order diffracted beams. For example, in a previous work showing electron diffraction from an optical standing wave, even in the Bragg regime of (+1)-order diffraction, the (-1)-order and (+2)-order diffracted beams are clearly visible in both theory and experiment, with the peak diffraction intensity of the (-1)-order half as high as that of the (+1)-order [17]. For nanofabricated gratings, diffraction efficiency can be improved by moving from amplitude gratings (e.g. a grating made from a 1-µm-thick platinum foil) to phase gratings (e.g. a grating made from a 30-nm-thick silicon nitride membrane), and carefully controlling the surface profile of the grating [12]. All these efforts improved the efficiency of various types of electron beam splitters, but intensity loss has never been completely eliminated.

Quantum mechanical "interaction-free measurement" (IFM) was proposed by Elitzur and Vaidman as a means of detecting the presence of an object without interacting with it [18]. In the concept of quantum IFM, a single probe-particle, such as a single photon, is sent to a Mach-Zehnder interferometer. The object to be detected is placed in one of the two paths of the interferometer and fully blocks the probe-particle if the particle hits the object. The presence of the object can change the output state of the interferometer, and it is possible in principle to detect the object without interaction between the probe-particle and the object. The quantum IFM concept has been further developed by incorporating the quantum Zeno effect [19]. Efficient quantum IFM was realized by cascading multiple stages of Mach-Zehnder interferometers and using highly asymmetric beam splitters. The efficiency of quantum IFM can approach unity by repeated interrogations of the object with a small fraction of the probe-particle wavefunction, while the interaction probability between the probe-particle and the object simultaneously tends to zero. While early development of quantum IFM chose photons as the probe-particles, recent works have proposed to perform quantum IFM with electrons, which holds potential to reduce sample damage in electron microscopy [3,15]. Here, we use the concept of quantum IFM incorporated with quantum Zeno effect to suppress spurious coupling to undesired modes in a quantum mechanical system (in our case, an electron beam splitter). We call this quantum "interaction-free suppression" (IFS).

In this paper, we propose a highly efficient two-port electron beam splitter that utilizes quantum IFS. The theoretical efficiency can be made arbitrarily close to unity. The beam splitter consists of a weak phase grating, such as a nano-patterned membrane, and a resonator. Beam-splitting is achieved by passing the electron beam through the weak phase grating multiple times within the resonator. The beam-splitting ratio is controlled by the number of passes through the grating. Higher-order diffraction can be suppressed by inserting an aperture that simply blocks unwanted diffracted beams. The loss introduced by this aperture can be arbitrarily close to zero according to quantum IFS.

The paper is organized as follows. In Sec. II, we introduce the scattering matrix method for theoretical calculation, and analyze the working principle of the beam splitter under the simplest scenario – the two-beam condition – by considering only two beams. In Sec. III, we analyze the beam splitter by taking into account high-order diffracted beams, and propose to reduce the intensity loss due to high-order beams with a beam-blocking aperture. In Sec. IV, we discuss the effect of electron inelastic scattering. In Sec. V, we show the aperture performs quantum IFM on the electron, and evaluate the total intensity loss of the beam splitter. Finally, Sec. VI summarizes our results.

## II. TWO-BEAM CONDITION OF A CRYSTAL BEAM SPLITTER: SCATTERING MATRIX METHOD

Our beam splitter design consists of a weak phase grating and a resonator. Figure 1(a) shows the schematic of the design. The beam splitter is a two-port system with one input port and two output ports. The input and output ports have gates that control the entrance and exit of the electron. The beam splitter works with a pulsed electron beam. The incident electron enters the resonator through the input port gate, the gate is then closed, and the electron starts to bounce back and forth in the resonator. A weak phase grating is placed in the resonator. We assume, for the purpose of explaining the basic operation of the beam splitter, that the grating imposes a pure phase modulation onto the electron beam without amplitude modulation caused by inelastic scattering. The electron goes through the grating multiple times when it is resonating, and is diffracted by the grating. For now, we assume there is only one diffracted beam. The electron is in the direct beam with unity probability prior to its first interaction with the diffraction grating. After each diffraction event, a small fraction of the electron wavefunction is diffracted (the fraction is small as a weak phase grating is used). Hence, as the electron passes through the grating multiple times, the probability of the electron occupying the diffracted beam builds up coherently. The beam-splitting ratio, i.e. the relative intensity of the direct and diffracted beams, depends on the number of passes through the grating. If the output gates open after a certain number of passes and the direct and diffracted beams go to output ports 1 and 2 respectively, a corresponding splitting ratio can be achieved by the beam splitter. Here we want to mention the resonator mirror forms an image with -1 magnification. In order to impose the same phase modulation to both forward-going and backward-going electron beams, the grating should be symmetric with respect to the optical axis. Otherwise, one would need a beam deflection or rotation device to make sure the same phase modulation is imposed every time the electron passes through the grating. Furthermore, the complete design involves several other electron-optics components, such as electron sources for the input beam, electron detectors for the output beams, and lenses for transforming between focused beams at the input and output ports and plane waves illuminating the weak phase grating. However, we focus on the fundamental operating principles and leave out details of the electron source, lenses, deflectors, and detectors, as shown in Fig. 1(a).

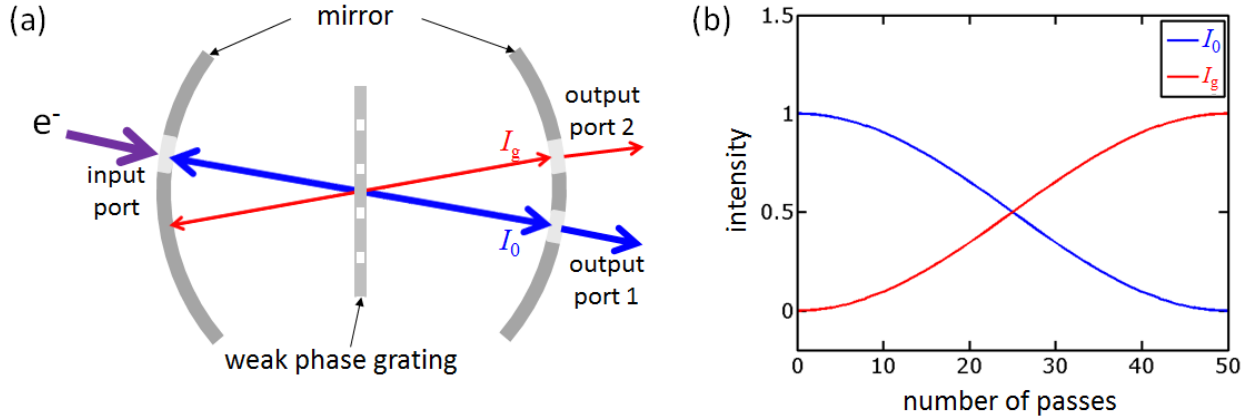

FIG. 1. (a) Schematic of the beam splitter design. The two-port beam splitter has one input port and two output ports. The input electron enters a resonator in which a weak phase grating is placed. The electron is diffracted by the grating. After a certain number of round-trips (and passes through the grating in the resonator), the electron leaves the resonator. The output ports 1 and 2 corresponds to the direct (blue) and diffracted (red) beams, respectively. In the schematic, we focus on the fundamental operating principles and leave out details of the electron source, lenses, deflectors, and detectors. (b) Calculated intensities of direct and diffracted beams as a function of number of passes through a crystal beam splitter, of which the thickness is 1% of the extinction distance. Beam-splitting ratio, i.e. the relative intensity between the two output beams, can be tuned by changing the number of passes through the crystal.

We adopt a scattering matrix method to analyze the beam splitter. We first introduce the method for a thin crystal as the weak phase grating in a two-beam condition, as shown in Fig. 1(a). The crystal is tilted with respect to the incident beam, so that only one diffracted beam is strongly excited. We model the direct beam and the diffracted beam in free-space as two plane waves with different momentum vectors: $e^{2\pi i k_0 \cdot r}$ (the direct beam) and $e^{2\pi i k_g \cdot r}$ (the diffracted beam). Before entering the crystal, the electron wavefunction is a superposition of the two plane waves:

$$\Psi_{in} = a_1 e^{2\pi i k_0 \cdot r} + a_2 e^{2\pi i k_g \cdot r}. \tag{1}$$

After exiting the crystal, the electron wavefunction is also a superposition of the plane waves:

$$\Psi_{out} = b_1 e^{2\pi i k_0 \cdot r} + b_2 e^{2\pi i k_g \cdot r}. \tag{2}$$

The crystal diffraction modulates the amplitudes of the two plane waves, and can be mathematically modeled as a scattering matrix $S$ in the following input-output relation

$$\begin{bmatrix} b_1 \\ b_2 \end{bmatrix} = S \begin{bmatrix} a_1 \\ a_2 \end{bmatrix}, \tag{3}$$

which relates the amplitudes of the direct and diffracted beams at the input and output of the crystal. The scattering matrix $S$ for a thin crystal is shown to be [20,21]

$$S = \begin{bmatrix} \cos(\pi t/\xi_g) & i \cdot \sin(\pi t/\xi_g) \\ i \cdot \sin(\pi t/\xi_g) & \cos(\pi t/\xi_g) \end{bmatrix}, \tag{4}$$

where $\xi_g$ is the crystal extinction distance for electron diffraction. The scattering matrix $S$ is unitary and the total intensity of the electron beam is conserved, as expected. As described previously, the electron beam will pass through the crystal multiple times, hence

$$\Psi_{OUTPUT} = S^N \Psi_{INPUT}, \tag{5}$$

with $\boldsymbol{\Psi}$ being the vector representation of the electron wavefunction where the first (second) element is the amplitude coefficient of plane wave $e^{2\pi i \boldsymbol{k_0} \cdot \boldsymbol{r}}$ ($e^{2\pi i \boldsymbol{k_g} \cdot \boldsymbol{r}}$).

Assume the electron starts in the direct beam, namely,

$$\boldsymbol{\Psi}_{\text{INPUT}} = \begin{bmatrix} 1 \\ 0 \end{bmatrix}, \tag{6}$$

we can calculate the final intensities of the direct and diffracted beams after the electron passes through the beam splitter $N$ times,

$$I_0 = |\boldsymbol{\Psi}_{\text{OUTPUT}}(1)|^2, \tag{7}$$

$$I_g = |\boldsymbol{\Psi}_{\text{OUTPUT}}(2)|^2, \tag{8}$$

where the index 1 (2) is referring to the first (second) element of the wavefunction vector. The intensities are calculated for a crystal with a thickness $t$ that is 1% of the extinction distance $\xi_g$, and the results are shown in Fig. 1(b). The incident electron is originally in the direct beam. As the electron passes through the crystal, a small fraction of the beam intensity will be split into the diffracted beam via electron diffraction. With an increasing number of passes through the crystal, intensity in the diffracted beam will coherently build up. When a certain number of passes is reached ($N = 50$ in Fig. 1(b)), the beam intensity is completely transferred from the direct beam to the diffracted beam. We call this number of passes the switch point. This switch point depends on the phase modulation of the weak phase grating. The weaker the phase modulation of the grating is, the larger the switch point. In this example where the weak phase grating is a thin crystal, changing the phase modulation can be achieved by varying the crystal thickness $t$, thus varying the $\pi t/\xi_g$ term in the scattering matrix. As a result, if the output port gates are opened after the electron passes through the crystal a certain number of times, a corresponding splitting ratio between output port 1 (the direct beam) and output port 2 (the diffracted beam) can be achieved. This splitting ratio is tunable between zero and unity by controlling the number of passes. Thus, it can be seen that our electron beam splitter works very similarly to a microwave or photonic directional coupler [22,23].

**III. MULTI-BEAM CONDITION OF A NANOFABRICATED GRATING BEAM SPLITTER: BLOCKING HIGH-ORDER DIFFRACTION**

Electron diffraction from a periodic structure (a crystal or a nanofabricated grating) will inevitably produce multiple diffraction orders, which render the analysis in the two-beam condition questionable. Here, we consider a beam splitter configuration (Fig. 2(a)) similar to the design discussed in the previous section. This time we choose a nanofabricated grating as the weak phase grating, and include high-order diffraction in our analyses.

The scattering matrix method is again used to analyze the beam splitter, while taking high-order diffracted beams into consideration. Due to the existence of high-order diffraction, the dimension of the scattering matrix is now larger than two. To include all possible diffraction orders, $M$, the matrix dimension, would be infinite. In our calculation, we choose a sufficiently large number for $M$ so that a finite dimensional scattering matrix can still give accurate results (we chose $M = 100$). This is because beam intensities for very-high-order diffraction beams are weak and thus negligible.

We assume the weak phase grating is a one-dimensional sinusoidal phase grating; namely, the grating modulation of the transmission has the following form (for one period)

$$g_0(x) = exp\left[i\frac{A}{2}sin\left(2\pi\frac{x}{P}\right)\right], |x| \leq \frac{P}{2}. \tag{9}$$

Here $A$ is the phase amplitude of the grating, $P$ is the grating pitch, and the periodic profile is along the $x$-axis. Hence, the transmission modulation function of the full grating is the convolution between $g_0(x)$ and a delta pulse train

$$g(x) = g_0(x) * \sum_n \delta(x - nP). \tag{10}$$

This function can be cast into a Fourier series

$$g(x) = \sum_{n=-\infty}^{\infty} J_n\left(\frac{A}{2}\right) exp\left(i\frac{2\pi n}{P}x\right). \tag{11}$$

For periodic, sinusoidal phase modulation, the Fourier series coefficients are Bessel functions of the 1$^{st}$ kind. Therefore, the scattering matrix $S$ describing the weak sinusoidal phase grating is a $(2M + 1) \times (2M + 1)$ matrix with elements

$$S_{ij} = J_{(j-i)}\left(\frac{A}{2}\right). \tag{12}$$

Note that if the order $(j - i)$ is odd and negative, the Bessel function $J_{(j-i)}\left(\frac{A}{2}\right)$ is negative (for small phase amplitude $A$). Again, if $\Psi_{INPUT}$ and $\Psi_{OUTPUT}$ are $(2M + 1)$-dimensional vectors representing the input and output electron wavefunctions including all diffraction orders (from $(-M)$-order to $M$-order), the input-output relation is

$$\Psi_{OUTPUT} = S^N \Psi_{INPUT}, \tag{13}$$

for an electron passing through the grating $N$ times. The intensity of each diffraction order can be obtained by taking the square of the magnitude of the corresponding element in the vector representation of the wavefunctions. The intensities were calculated for a beam splitter using a weak sinusoidal phase grating with a phase amplitude $A = 0.02\pi$ (assume electron energy is 200 keV, this phase modulation can be achieved by a nanofabricated grating made from a 1 nm thick amorphous carbon film), and the results for 0-th and (+1)-st order beams are shown in Fig. 2(b). These are the two beams exiting from the output ports. Intensity transfer between the direct (0-th order) and diffracted ((+1)-st order) beams is still observed. However, in contrast to the two-beam condition, when the intensity of the direct beam drops to zero, the intensity of the diffracted beam is not unity; namely, a complete intensity transfer between the direct and diffracted beams cannot be achieved at the switch point. Moreover, the sum of the intensities of the two output beams is always below unity for any number of passes greater than zero. If used as a two-port beam splitter, this less-than-unity intensity sum means imperfect efficiency, or intensity loss, associated with the beam splitter. The intensity loss is due to the existence of high-order diffraction. There is a probability that the electron is in a high-order diffracted beam rather than the direct or (+1)-st order diffracted beam, hence the intensity sum of the two output beams is less than unity.

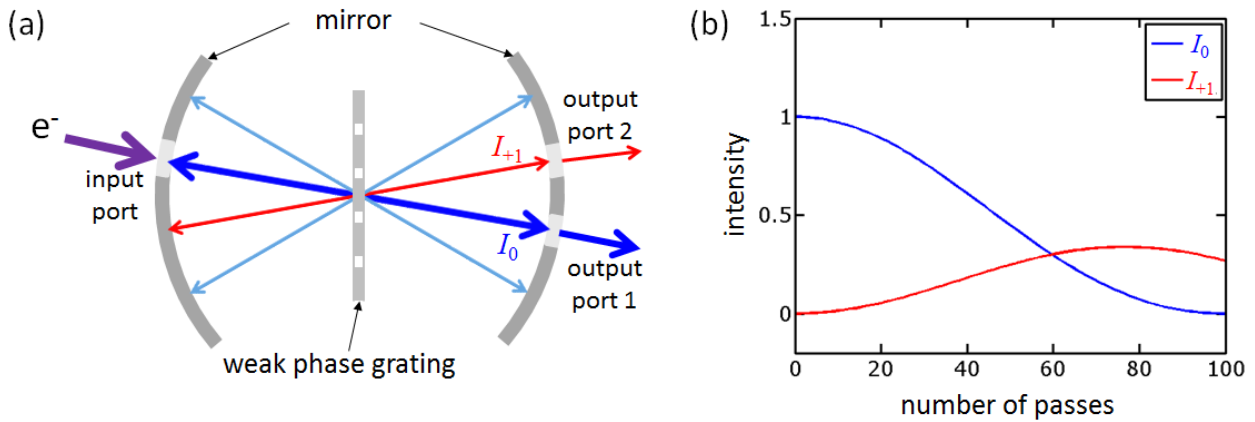

FIG. 2. (a) Schematic of the beam splitter design considering high-order diffraction. The two-port beam splitter has one input port and two output ports. The input electron enters a resonator in which a weak phase grating is placed. The electron is diffracted by the grating. After a certain number of round-trips and passes through the grating in the resonator, the electron leaves the resonator. The output ports 1 and 2 corresponds to the direct (blue) and diffracted (red) beams, respectively. High-order diffracted beams are also shown (light blue), but they do not contribute to the output beams. (b) Calculated intensities of direct (0-th order) and diffracted ((+1)-st order) beams as a function of number of passes through a weak sinusoidal phase grating, of which the phase amplitude is $0.02\pi$. Beam-splitting ratio, i.e. the relative intensity between the two output beams, can be tuned by changing the number of passes through the grating.

We use quantum IFS to reduce the intensity loss of the beam splitter. The idea is shown schematically in Fig. 3(a). Compared to Fig. 2(a), only one additional component, a beam-limiting aperture, is added. This aperture is placed at a plane, in which the high-order diffracted beams are clearly separated from the direct (0-th order) and diffracted ((+1)-st order) beams. This could be achieved by using a lens that transforms the diffracted beams into diffraction spots in the back focal plane, and place the aperture at that plane. The aperture allows the direct and diffracted beams to pass through, while completely blocking high-order diffraction. Each time the electron passes through the grating, it will be diffracted and a fraction of the intensity will go to high-order diffraction. However, the presence of the aperture within the resonator acts on high-order diffraction in every round trip and prevents intensity build-up in high-order diffraction.

We use the scattering matrix method similar to the abovementioned procedure to analyze the beam splitter with high-order diffraction and a limiting aperture. The scattering matrix $S$ associated with the weak phase grating will remain the same. Operation of the limiting aperture can be represented by the following scattering matrix

$$S_{\text{aper}} = diag(\cdots, 0, 1, 1, 0, \cdots). \tag{14}$$

This is a diagonal matrix with only two nonzero elements corresponding to the direct and diffracted beams that are not blocked by the aperture. The total effective scattering matrix is the multiplication of $S$ and $S_{\text{aper}}$, so that the input-output relation becomes

$$\Psi_{\text{OUTPUT}} = (SS_{\text{aper}})^N \Psi_{\text{INPUT}}, \tag{15}$$

for an electron passing through the grating $N$ times. Again, the beam intensities were calculated for a beam splitter using a weak sinusoidal phase grating with a phase amplitude $A = 0.02\pi$, and the results for 0-th and (+1)-st order beams are shown in Fig. 3(b). Unlike the case where high-order diffraction is not blocked, the beam splitter with a limiting aperture achieves almost complete intensity transfer between the direct and diffracted beams at the switch point, and the sum of the intensities of the two output beams is near unity. This unity intensity sum is almost the same as in the two-beam condition (Fig. 1(b)), even though higher-order diffracted beams are included. Despite the existence of high-order diffracted beams, the limiting aperture prevents intensity build-up in these beams, thus minimizing intensity loss. With minimal intensity loss, a highly efficient, two-port electron beam splitter based on electron diffraction from a nanofabricated grating or a crystal can be achieved.

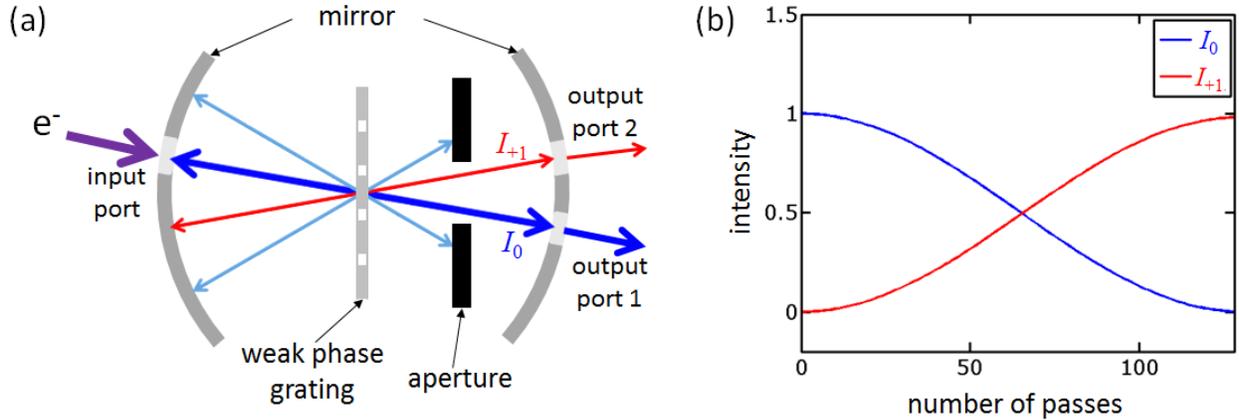

FIG. 3. (a) Schematic of the beam splitter design using quantum IFS effect to remove intensity loss in high-order diffraction. The input electron enters a resonator in which a weak phase grating is placed. The electron is diffracted by the grating. After a certain number of round-trips and passes through the grating in the resonator, the electron leaves the resonator. The output ports 1 and 2 corresponds to the direct (blue) and diffracted (red) beams, respectively. High-order diffracted beams are also shown (light blue). A limiting aperture (black) placed within the resonator allows the direct and diffracted beams to pass through, while completely blocking high-order diffraction. (b) Calculated intensities of direct (0-th order) and diffracted ((+1)-st order) beams as a function of number of passes through a weak sinusoidal phase grating, of which the phase amplitude is $0.02\pi$, followed by a limiting aperture. Beam-splitting ratio, i.e. the relative intensity between the two output beams, can be tuned by changing the number of passes through the grating.

## IV. EFFECT OF INELASTIC SCATTERING

In the beam splitter design, loss can be caused by not only the undesired diffraction modes, but also inelastic scattering when the electron passes through the weak phase grating. The inelastic intensity loss depends on the type of the weak phase grating. In this section, we focus on the situation where a nanofabricated grating is used as the weak phase grating. We consider the schematic showing in Fig. 3(a), with a nanofabricated grating as the weak phase grating. This time, the grating imposes both phase and amplitude modulations onto the electron beam, with the amplitude modulation introduced by inelastic scattering. We model the inelastic scattering by assigning the following intensity transmission probability when the electron passes through the grating:

$$T = e^{-t/\lambda}. \tag{16}$$

Here, $t$ is the material thickness of the grating, and $\lambda$ is the inelastic scattering mean free path (MFP) of the material. We modified the scattering matrix method by considering this transmission probability, and calculated beam intensities for a beam splitter using a weak sinusoidal phase grating with a phase amplitude $A = 0.02\pi$. For 200 keV electron energy, we chose 1-nm-thick amorphous carbon film as the grating material, which has an inelastic MFP of 160 nm [24]. Figure 4(a) shows the calculated 0-th order and (+1)-st order beam intensities. As expected, inelastic scattering leads to a non-ideal efficiency, with the sum of 0-th order and (+1)-st order beam intensities below unity. At the switch point, the efficiency is ~55%. To investigate the effect of grating materials, we also considered another nanofabricated grating made from 1-nm-thick gold foil. For 200 keV electron energy, this grating imposes a phase modulation with amplitude $A = 0.058\pi$, and inelastic MFP for gold is 84 nm. Figure 4(b) shows the calculated 0-th order and (+1)-st order beam intensities for this grating. Similar to Fig. 4(a), intensity loss is caused by inelastic scattering. At the switch point, the efficiency is ~63%. It can be seen that the choice of the grating material has an effect on the beam splitter efficiency. For a nanofabricated grating, a periodically structured material presents a periodic potential to the electron beam, causing electron diffraction and beam splitting. This potential is the mean inner potential (MIP) of the material. To achieve a certain beam-splitting ratio, the grating needs to impose a certain phase shift $\Delta\phi_0$ to the electron beam, which requires a certain material thickness $t_0$ according to the following equation:

$$\Delta\phi_0 = C_0 V_{\text{MIP}} t_0. \tag{17}$$

Here, $V_{\text{MIP}}$ is the material MIP, and $C_0$ is a constant. The beam splitter efficiency can be estimated by the intensity transmission probability

$$T_0 = e^{-t_0/\lambda} = e^{-\frac{\Delta\phi_0}{C_0 V_{\text{MIP}} \lambda}}. \tag{18}$$

Hence, materials with a large MIP ($V_{\text{MIP}}$) and a large MFP ($\lambda$) are preferred for a high efficiency beam splitter. Figure 4(c) surveys reported MIP and MFP values for several materials [24–29]. Different data points on the MIP-MFP plot represent different materials with the corresponding MIP and MFP values. Gratings made from materials with the same MIP-MFP product ($V_{\text{MIP}}\lambda$) should lead to a similar beam splitter efficiency. The three red dashed lines in Fig. 4(c) indicate three contours of constant $V_{\text{MIP}}\lambda$ value: 1200 nm · V, 1600 nm · V, and 2000 nm · V.

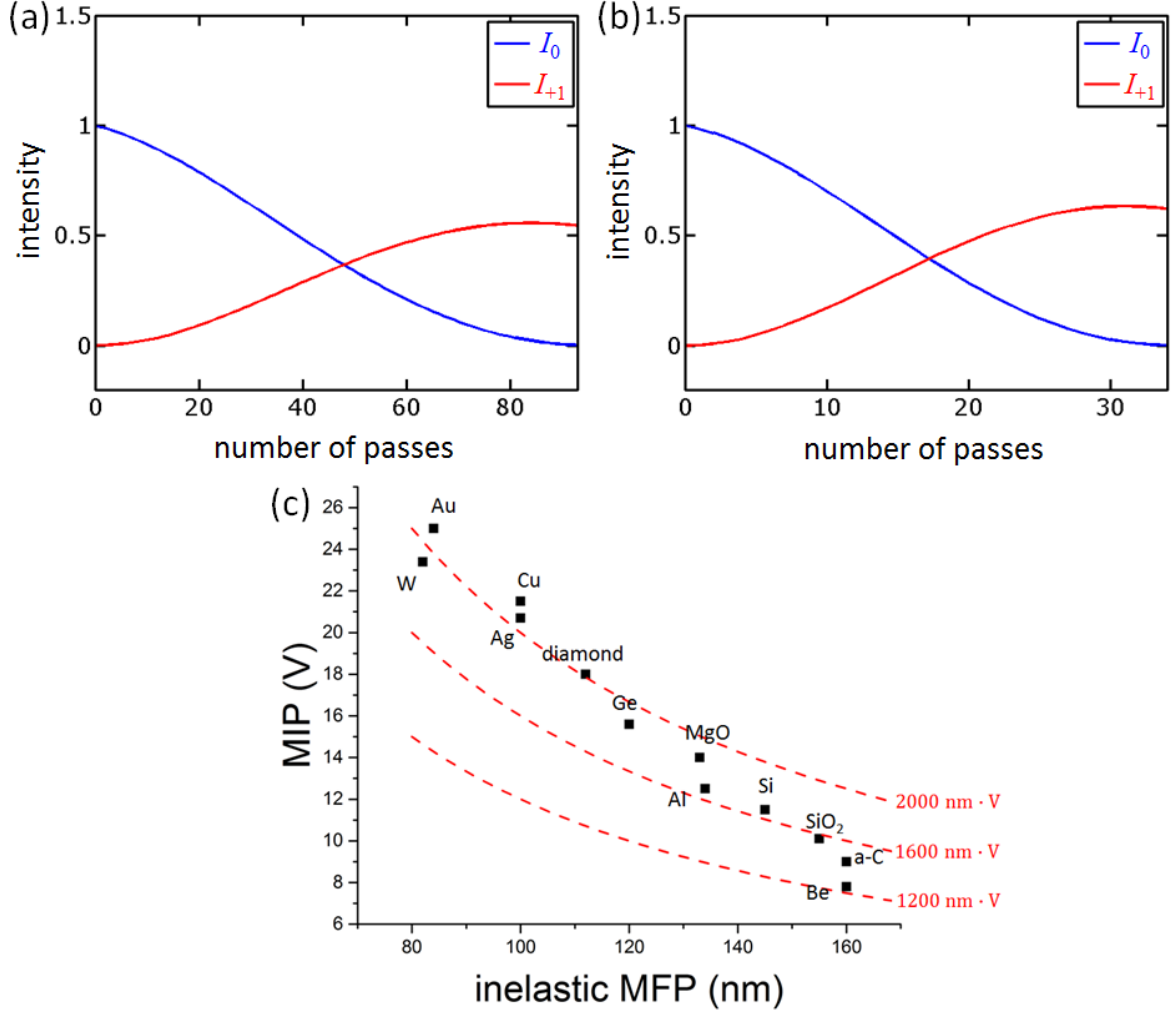

FIG. 4. The effect of inelastic scattering on the performance of the beam splitter. (a) Calculated intensities of direct (0-th order) and diffracted ((+1)-st order) beams as a function of number of passes through a weak sinusoidal phase grating made from a 1-nm-thick amorphous carbon film. Electron energy is 200 keV. Phase amplitude of the grating is $0.02\pi$, and electron mean free path is 160 nm. Quantum IFS is used to suppress high-order diffraction. At the switch point, the efficiency is ~55%. (b) Calculated intensities of direct (0-th order) and diffracted ((+1)-st order) beams as a function of number of passes through a weak sinusoidal phase grating made from a 1-nm-thick gold foil. Electron energy is 200 keV. Phase amplitude of the grating is $0.058\pi$, and electron

mean free path is 84 nm. Quantum IFS is used to suppress high-order diffraction. At the switch point, the efficiency is ~63%. (c) Reported MIP and MFP values for several materials [24–29]. Red dashed lines are contours with constant MIP-MFP product ($V_{\text{MIP}}\lambda$).

The above analysis on inelastic scattering applies to the beam splitter design with a nanofabricated grating as the weak phase grating. The calculated efficiency also applies to this specific design only. Similar analysis can be performed for other types of weak phase gratings, such as crystalline materials and optical standing waves. We will not present detailed analysis here as it is beyond the scope of this paper.

## V. DISCUSSION

It may appear paradoxical that intensity loss due to higher-order diffracted beams can be reduced simply by blocking these beams with an aperture. When a beam is blocked, its intensity, or the probability that the electron is in this beam, is lost permanently. Therefore, one would naively expect the loss to remain the same, if not increase, when high-order diffraction is blocked. However, the combination of quantum IFM with the quantum Zeno effect provides a counterintuitive route to reduce loss due to high-order diffraction. In quantum IFM, a perfectly opaque object can be detected by a probe-particle without losing the particle, as the probe-particle repeatedly interrogates the object with a small fraction of the particle wavefunction obtained via an asymmetric beam splitter. To draw an analogy, the efficient, two-port electron beam splitter proposed here performs a quantum IFM: the electron is analogous to the probe-particle; the weak phase grating acts as the asymmetric beam splitter; the high-order diffraction is analogous to the small fraction of the particle; the limiting aperture is the opaque object; and the resonator effects the repeated interrogation. As a result, intensity loss due to high-order diffraction is eliminated, as with the loss of the particle in quantum IFM.

Here, we want to point out that after introducing the limiting aperture, the intensity loss due to high-order diffraction is not exactly zero, although its value can be made arbitrarily low. This finite intensity loss can be observed in Fig. 3(b). At the switch point, the direct beam intensity drops to zero, and the diffracted beam intensity approaches but never achieves unity. Hence, there is a finite intensity loss. We calculated this intensity loss at the switch point for beam splitter designs with different switch points. The results shown in Fig. 5 demonstrate that intensity loss drops with increasing switch point. Below 1% intensity loss due to high-order diffraction can be achieved for a switch point greater than ~230. As mentioned before, the switch point can be increased by reducing the phase modulation of the grating. Theoretically there is no upper limit to the switch point. As a result, the beam splitter can be designed so that arbitrarily low intensity is lost. In practice, the switch point could be limited by the weak phase grating. If a nanofabricated grating is used as the weak phase grating, the minimum achievable phase modulation can be limited by the electron energy and the minimum achievable material thickness, thus imposing an upper limit to the switch point. For instance, for an electron with 60 keV energy, a nanofabricated grating made from monolayer graphene imposes the minimum achievable phase modulation, which is ~$0.01\pi$ (considering graphene as a uniform film with atomic scale thickness). This phase modulation leads to the maximum achievable switch point, which is 193, at this electron energy. However, if an optical standing wave is used as the weak phase grating, a weak phase modulation can be achieved by lowering the light intensity, and there is no upper limit to the switch point. For instance, a laser beam with 1064 nm wavelength and 5 mm beam waist is used to make the optical standing wave. It forms a Kapitza-Dirac diffraction grating for electrons with $2\times10^6$ m/s velocity or 12 eV energy. To obtain a similar phase modulation as above (0.01π), the required laser intensity is estimated to be about 3 GW/m². If the laser intensity is reduced, a smaller phase modulation, and a larger switch point, can be achieved. For Kapitza-Dirac diffraction, in theory there is no upper limit to the switch point, and hence the intensity loss due to high-order diffraction can be made arbitrarily small.

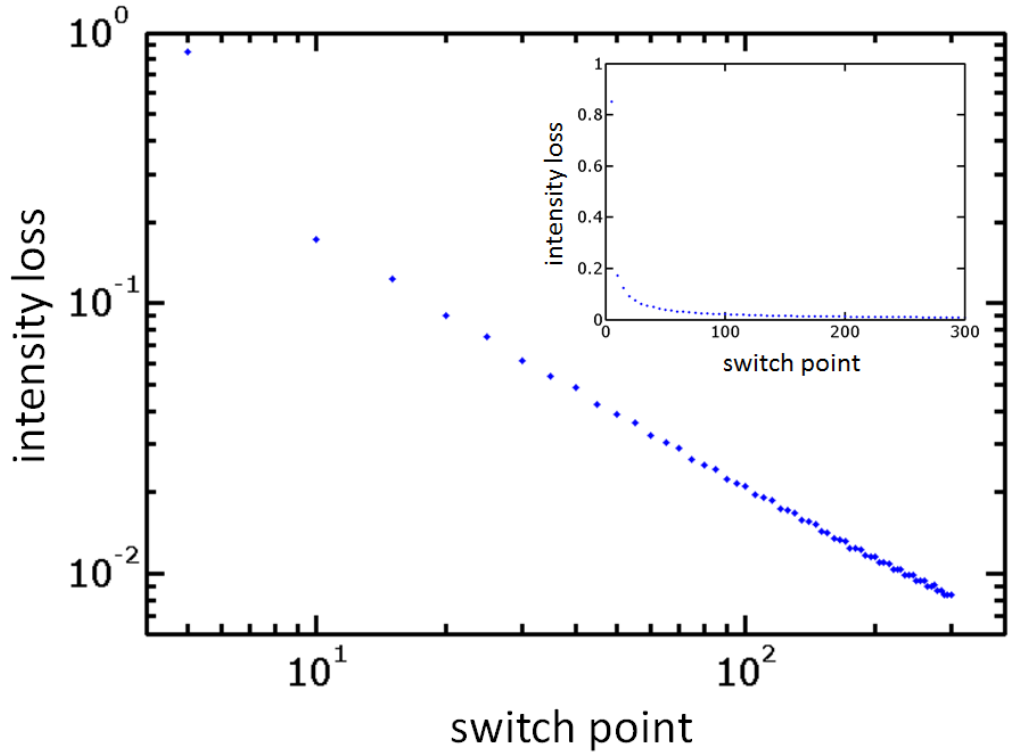

FIG. 5. Beam splitter intensity loss at the switch point for designs with different switch points (log-log scale). Inelastic scattering is neglected in this calculation. The intensity loss decreases monotonically with increasing switch point and approaches zero. Inset: the same plot with a linear scale.

Despite the fact that intensity loss due to high-order diffraction can be made arbitrarily close to zero, there is still finite intensity loss caused by inelastic scattering when the electron passes through the weak phase grating. Inelastic scattering is a non-unitary evolution described by a non-Hermitian Hamiltonian, and hence cannot be reduced by a combination of quantum IFM with quantum Zeno effect that successfully suppresses the loss due to high-order diffraction [30,31]. It should be noted that this intensity loss is independent of the choice of switch point. To minimize the impact of inelastic scattering, higher electron energy and materials with lower inelastic scattering cross sections are preferred. Alternatively, choosing a different implementation of the weak phase grating (e.g. using optical standing waves) can also reduce inelastic scattering.

To experimentally implement the proposed device, several practical issues need further consideration. Firstly, the electron-optical system should have a good alignment. In the proposed beam splitter, the electron beam passes through the grating multiple times. For each pass, the phase modulation of the electron beam has to be aligned with the grating structure, so that the phase modulation can be enhanced by passing through the grating. In experimental implementation, misalignment should be kept below a tiny fraction of the grating period, so that alignment error is small even after multiple passes. Secondly, the proposed device works with pulsed electron beams by using switchable gates. These gates can switch between an aperture state that passes the electron into and out of the resonator, and a mirror state that reflects the electron beams. Such an electron-optical component requires further development. Thirdly, the requirement on the spatial and temporal coherence of the electron beam, as well as aberrations introduced by the electron-optical components, need future investigation.

## VI. CONCLUSION

We present a design for a highly efficient, two-port electron beam splitter. The beam splitter consists of a resonator and a weak phase grating. The input electron enters the resonator and bounces back and forth while passing through the weak phase grating multiple times. Depending on the targeted beam-splitting ratio, the electron exits via the two output ports after some number of round-trips. We demonstrated a scattering matrix method to analyze the performance of the beam splitter, and showed its working principle in the two-beam condition. However, we found that the efficiency of the beam splitter was compromised, because higher diffraction orders were generated when the electron passed through the weak phase grating. This issue can be solved by introducing a limiting aperture that fully blocks the high-order diffraction in each round-trip. This technique utilizes quantum interaction-free measurement, and intensity loss can be made arbitrarily low, or equivalently, the efficiency of the beam splitter can be made arbitrarily close to unity. In practice, nearly all components (electron sources, mirrors, phase gratings, apertures, and detectors) required to build such a beam splitter are well developed. Meanwhile, a gate that has two states, a mirror state and an aperture state, is under active development [3]. Therefore, experimental implementation of the beam splitter is feasible. This efficient, two-port beam splitter can find various applications in electron beam technologies, especially the emerging methods that bear a great resemblance to quantum optics experiments [3,15].

Additionally, our beam splitter design can be used to generate electron vortex beams with high efficiency. An electron vortex beam is a free electron beam that carries an orbital angular momentum (OAM) $l\hbar$, with $l$ the so-called topological charge. The electron vortex beams have been generated mainly by passing an electron beam through a nanofabricated diffraction hologram [11]. The pattern of the hologram is a grating with a fork defect, as illustrated in Fig. 6(a). Diffracted beams from this hologram carry different OAM depending on the diffraction order. Our beam splitter design enables the selection of one electron vortex beam carrying a specific OAM, without suffering intensity loss in other diffracted beams. Figure 6(b) shows an example of selecting an electron vortex beam with topological charge $l=2$. In our beam splitter design, if the weak phase grating is replaced by a weak phase hologram with the pattern in Fig. 6(a), each diffracted beam will carry an OAM with its topological charge same as its diffraction order. To select the vortex beam with topological charge $l=2$, the aperture placed in the resonator lets through the direct beam and the $2^{nd}$-order diffracted beam, while blocking all other beams. Quantum IFS ensures that, with the right number of passes through the hologram, beam intensity can be concentrated in the $2^{nd}$-order diffracted beam, with minimal intensity loss in the direct beam and other diffracted vortex beams. Hence, our scheme can generate a clean electron vortex beam carrying one specific OAM, like transforming a plane wave beam to a vortex beam, without sacrificing the beam intensity. Here we want to mention that this scheme requires a pulsed electron source with high transverse coherence. Previous reports demonstrated highly coherent pulsed electron sources for imaging and diffraction [32–34]. To get a fully coherent vortex beam, we would need a high degree-of-coherence, namely a transverse coherence length on the scale of the beam diameter.

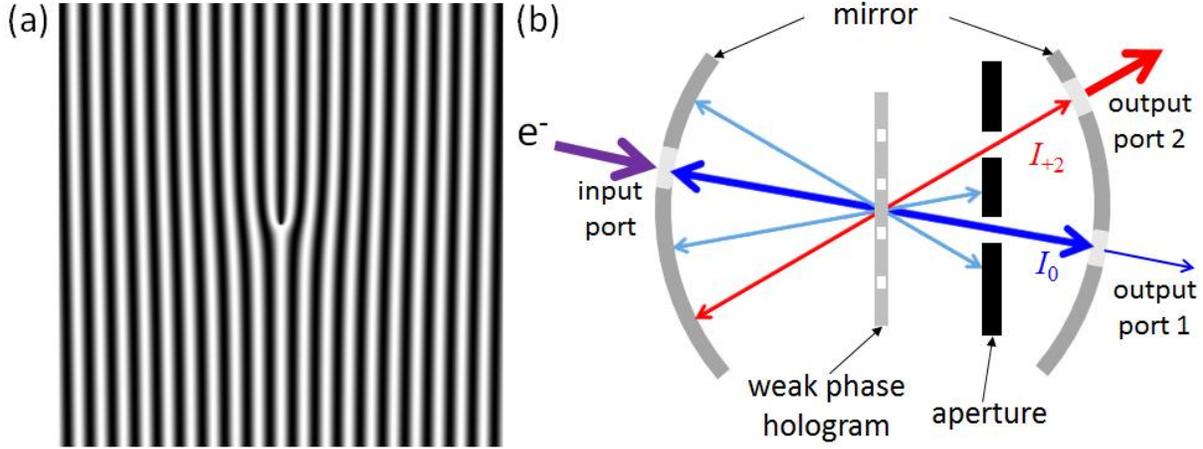

FIG. 6. Efficient generation of an electron vortex beam with a specific orbital angular momentum. (a) Diffraction hologram for vortex beam generation. The pattern is a grating with a fork defect. (b) Generation of an electron vortex beam with topological charge $l=2$. The weak phase hologram uses the pattern in (a). The aperture lets through the direct beam (blue) and the $2^{nd}$-order diffracted beam (red), while blocking all other beams (light blue).

Finally, we want to emphasize that this design can be applied to beam splitters for not only electrons, but also photons, neutrons, atoms, and any other quantum mechanical systems. A simple diffraction phase grating, combined with a resonator and an aperture, can be used to build two-port beam splitters for these systems with high efficiency. This design is effectively a method to convert a multiport system to a two-port system with minimal loss.

## ACKNOWLEDGEMENT

We gratefully acknowledge helpful discussions with J. Francis, T. Juffmann, C. Kealhofer, B. Klopfer, C. Kohstall, G. Skulason, and M. Kasevich from Stanford University, and J. Hoffrogge, S. Thomas, J. Hammer, D. Ehberger, S. Heinrich, P. Weber, and P. Hommelhoff from Friedrich Alexander University Erlangen-Nürnberg. This work was supported by Gordon & Betty Moore Foundation.

Ultramicroscopy **188**, 85 (2018).